\documentclass[letterpaper, 10 pt, conference]{ieeeconf}  % Comment this line out if you need a4paper

\IEEEoverridecommandlockouts                              % This command is only needed if
\usepackage{tikz}
\usetikzlibrary{calc}

\usepackage{graphicx}
\usepackage{epstopdf}
\usepackage{amsmath}
\usepackage{multicol}
\usepackage{stfloats}
\usepackage{amsfonts}
\usepackage{mathrsfs}
\usepackage{color}
\usepackage[ruled]{algorithm2e}
\usepackage{fancybox}
\usepackage{subcaption}

\usepackage{framed}
\usepackage{psfrag, amssymb, amsmath, cite}
\usepackage[colorlinks,linkcolor=red,anchorcolor=blue,citecolor=blue]{hyperref}

\newtheorem{remark}{Remark}
\newtheorem{definition}{\textbf{Definition}}

\newtheorem{theorem}{\textbf{Theorem}}

\newtheorem{assumption}{\textbf{Assumption}}
\newtheorem{lemma}{\textbf{Lemma}}

\title{\LARGE \bf
Temporal viability regulation for control affine systems \\
with applications to mobile vehicle coordination \\ under time-varying motion constraints
}

\author{Marcus Greiff, Zhiyong Sun, Anders Robertsson and Rolf Johansson% <-this % stops a space
\thanks{*The research leading to these results has received funding from
the Swedish Science Foundation (SSF) project ``Semantic mapping
and visual navigation for smart robots'' (RIT15-0038). }% <-this % stops a space
\thanks{The authors are members of the LCCC Linnaeus Center and
the ELLIIT Excellence Center at Lund University, Sweden. Emails:
        {\tt\small \{marcus.greiff, zhiyong.sun, rolf.johansson, anders.robertsson\} @control.lth.se}}%
}

\begin{document}

\maketitle
\thispagestyle{empty}
\pagestyle{empty}

%%%%%%%%%%%%%%%%%%%%%%%%%%%%%%%%%%%%%%%%%%%%%%%%%%%%%%%%%%%%%%%%%%%%%%%%%%%%%%%%
\begin{abstract}
Controlled invariant set and viability regulation of dynamical control systems have played important roles in many control and coordination applications.
In this paper we develop a \textit{temporal} viability  regulation theory for general dynamical control systems, and in particular for control affine systems. The time-varying viable set is parameterized by time-varying constraint functions, with the aim to regulate  a dynamical control system to be invariant in the time-varying viable set so that temporal state-dependent constraints are enforced. We consider both time-varying equality and inequality constraints in defining a temporal viable set. We also present sufficient conditions for the existence of feasible control input for the control affine systems. The developed temporal viability regulation theory is applied to mobile vehicle coordination.

\end{abstract}

%%%%%%%%%%%%%%%%%%%%%%%%%%%%%%%%%%%%%%%%%%%%%%%%%%%%%%%%%%%%%%%%%%%%%%%%%%%%%%%%
\section{Introduction}

In practice, control engineering applications often involve various constraints to guarantee system performance or general control efficiency. For example, in autonomous control of robotic manipulators, constraints could be imposed in designing feasible control strategy to ensure collision avoidance, safe human operation, or optimal trajectory generation. In the context of mobile vehicle coordination, a control task usually includes many types of inter-vehicle constraints described by equality/inequality constraints in geometric variables. In general, a control system is often subject to state constraints that limit admissible control inputs which should regulate the possible state-trajectories of the system. Therefore, designing a control input for a dynamical control system that meets the performance requirement or safety guarantee described by certain constraint functions is often a priority before proceeding with a real-time implementation.

A useful tool in dealing with various state-dependent constraints for control systems is controlled invariant theory \cite{blanchini1999set}, which has relevance to viability theory \cite{aubin2009viability} for dynamical systems in general. A set is termed `controlled invariant' under a dynamical control system, if the states of the control system are regulated to stay in the set with feasible control actions. Controlled invariant sets, or viable sets, are often parameterized by certain  equality/inequality  functions to meet system specifications or performance requirements. When a constraint is about to be violated, corrective actions for the control system should be undertaken that lead to viable control inputs and thus constraints are to be enforced.

The idea of controlled invariant set and viability regulation for  dynamical control systems has been deeply explored in recent years with many insightful and promising applications. Examples include barrier  verification of nonlinear and hybrid systems  \cite{prajna2007convex}, invariance regulation for safety control in robotic systems \cite{kimmel2016invariance,kimmel2017invariance},  obstacle avoidance and safety certificate  in vehicle navigation control \cite{panagou2016distributed,wang2017safety}, and feasible coordination for multiple mobile vehicle systems under motion constraints \cite{sun2018feasibility_short, sun2018feasibility}.

Control systems interacting with a dynamic environment will often involve \textit{time-varying} and state-dependent constraints, which demand time-varying viable functions in specifying temporal performance requirement. We remark that available results in the literature on controlled invariance and viability regulation mostly deal with time-invariant constraints or only state-dependent viable sets. Time-varying constraints have attracted some recent attention for some particular control systems, such as \cite{kimmel2016invariance, hauswirth2018time}. In this paper, we aim to develop general theories for time-varying viability regulation for general dynamical control systems, and in particular for control affine systems. We will present some conditions for designing a viable control input so that the solutions of a control affine system stay in a viable set defined by some time-varying equality/inequality functions of its state. Applications to mobile vehicle coordination control with time-varying motion constraints will be discussed as illustrative examples of the developed theory.

The main contributions of this paper include  a development of  temporal viability theory (with motivations of temporal contingent cone in a recent paper \cite{hauswirth2018time}), temporal viability regulation, and control law design for controlled temporal invariance of control affine systems. To illustrate their applications, we will consider two typical examples in mobile vehicle coordination control with time-varying motion constraints (in terms of distance and visibility maintenance), and justify their real-time performance guarantees with the developed temporal viability control.

This paper is organized as follows. Section \ref{sec:background} presents background on standard viability theory. Extensions to temporal viability theory are shown in Section \ref{sec:theory}. In Section \ref{sec:control_affine_system}, we focus on temporal viability regulation for control affine systems. Section \ref{sec:applications} provides certain typical application examples on mobile vehicle coordination under time-varying constraints. Conclusions in Section \ref{sec:conclusions} closes this paper.

\section{Background on  viability theory} \label{sec:background}
This section presents some background of controlled invariance and viability theory  from \cite{blanchini1999set,aubin2009viability}, which will motivate the development of temporal viability theory in the next section.

\begin{definition}
(\textbf{Viability and viable set}) Consider a control system in $\mathbb{R}^n$ described by a differential equation $\dot x(t) = f(x(t), u(t))$. A subset $\mathcal{F}\in \mathbb{R}^n$ enjoys the viability property for the system $\dot x(t)$  if for every initial state $x(0) \in \mathcal{F}$, there exists at least one solution to the system
starting at $x(0)$ which is viable in the time interval $[0, \bar t\,]$ in the sense that $$\forall t \in [0, \bar t \,], x(t) \in \mathcal{F}.$$
\end{definition}

%We assume the solution of the differential system $\dot x(t) = f(x(t), u(t))$, modeling vehicle control systems under constraints, is well defined. %Generalizations of the viability theory are also possible, by using the set-valued analysis and differential inclusion \cite{aubin2009set}.
%When a differential equation involves discontinuous right-hand side (e.g., switching controls), we understand its solutions in the sense of Filippov \cite{cortes2008discontinuous}.

Now define a distance function for a point $y$ to a set $\mathcal{F}$ as $d_\mathcal{F}(y)=:  \inf\limits_{z \in \mathcal{F}} \|y-z\|$, where $\|\cdot\|$ denotes the Euclidean 2-norm, and consider the definition of contingent cone as follows.
\begin{definition} \label{def:contingent_cone}
(\textbf{Contingent cone}) Let $\mathcal{F}$ be a nonempty subset of $\mathcal{X} \in \mathbb{R}^n$ and $x$ belongs to $\mathcal{F}$. The
contingent cone to $\mathcal{F}$ at $x$ is the set
\begin{align}
T_\mathcal{F}(x) = \left\{v \in \mathcal{X} |\,\,\,\,  \liminf\limits_{h \rightarrow 0^+} \frac{d_\mathcal{F}(x +hv)}{h} =0  \right\}. %_{h \right 0^+}  % \frac{d_K(x +hv)}{h} =0
\end{align}
\end{definition}
%It has been shown in \cite{blanchini1999set} that though the distance function $d_\mathcal{F}(y)$ depends on the considered norm, the set $T_\mathcal{F}(x)$ does not.
It is obvious that the set $T_\mathcal{F}(x)$ is non-trivial only on the boundary of $\mathcal{F}$. \footnote{It has been shown in \cite{blanchini1999set} that though the distance function $d_\mathcal{F}(y)$ depends on the considered norm, the set $T_\mathcal{F}(x)$ does not. }

A key result in the set-invariance analysis, the celebrated Nagumo theorem, is stated as follows (see \cite{blanchini1999set} or \cite{aubin2009viability}).
\begin{theorem}  \label{Theorem_Nagumo}
(\textbf{Nagumo theorem}) Consider the system
$\dot x (t) = f (x(t))$, and assume that, for each initial condition in
a set $\mathcal{X} \subset \mathbb{R}^n$, it admits a globally unique solution. Let $\mathcal{F} \subset \mathcal{X}$ be
a closed and convex set. Then the set $\mathcal{F}$ is positively
invariant for the system if and only if
\begin{align} \label{eq:Nagumo1}
f (x(t)) \in T_\mathcal{F}(x), \,\,\, \forall x \in \mathcal{F},
\end{align}
where $T_\mathcal{F}(x)$ denotes the \textit{contingent cone} of $\mathcal{F}$ at $x$.
\end{theorem}

Generalizations of the Nagumo theorem and viability theory are also possible, by using the set-valued analysis \cite{rockafellar2009variational} and differential inclusion \cite{aubin2009set}.
%For the definition of contingent cone, see \cite{aubin2009viability}.

If $x$ is an interior point in the set $\mathcal{F}$, then $T_\mathcal{F}(x) = \mathbb{R}^n$. Therefore, the condition in Theorem \ref{Theorem_Nagumo} is only meaningful when $x \in \text{bnd}(\mathcal{F})$, where $\text{bnd}(\mathcal{F})$ denotes the boundary of $\mathcal{F}$. Thus, the condition in \eqref{eq:Nagumo1} can be equivalently stated as
\begin{align}
f (x(t)) \in T_\mathcal{F}(x), \,\,\, \forall x \in \text{bnd}(\mathcal{F}).
\end{align}
The above condition clearly has an intuitive and geometric interpretation: if at $x \in \text{bnd}(\mathcal{F})$, the derivative
$\dot x   = f(x(t))$ points inside or  is tangent to $\mathcal{F}$, then the
trajectory $x(t)$ remains in $\mathcal{F}$.

% Now we consider a viable set $\mathcal{F}$ parameterized by an inequality associated with a $C^1$ (i.e., continuously differentiable) function $g(x): \mathbb{R}^n \rightarrow \mathbb{R}$,
% \begin{align} \label{eq:set_K}
% \mathcal{F} = \{x | g(x) \leq 0\}.
% \end{align}
% In this way, the calculation of $T_\mathcal{F}(x)$ is simplified to be
% \begin{align} \label{viable_condition_inequality}
% T_\mathcal{F}(x) = \{ v \in x | \left \langle v, \nabla g(x)\right \rangle \leq 0 \},
% \end{align}
% for any $ g(x) = 0$  and $T_\mathcal{F}(x) = \mathbb{R}^n$ when $g(x) < 0$. For the   set  $\mathcal{F}$ defined in \eqref{eq:set_K}, a consequence of Nagumo theorem is the following lemma on a controlled-invariant set.
% \begin{lemma} \label{lemma:invariant_set_control}
% (\textbf{Set-invariance in control}, \cite{blanchini1999set})  Consider a set $\mathcal{F}$ parameterized by an inequality of a continuously differentiable function $g(x)$: $\mathcal{F} = \{x| g(x) \leq 0\}$. Then the set $\mathcal{F}$ is positively invariant under the dynamic control system $\dot x(t) = f(x(t), u(t))$ if $\dot x(t) \in T_\mathcal{F}(x)$ of \eqref{viable_condition_inequality}, or equivalently
% \begin{align}
% \left< \nabla g(x),  f(x(t), u(t))  \right> \leq 0, \,\,\, \forall x:  g(x(t)) = 0.
% \end{align}
% \end{lemma}

\section{Theory of temporal viability}  \label{sec:theory}
In this section we present several general concepts on temporal viability and develop some novel results on controlled temporal invariance (termed `temporal viability regulation') for dynamical control systems.

Consider the following time-varying control dynamical system described by a general ordinary differential equation
\begin{align} \label{eq:control_system}
    \dot x(t) = f(x(t), u(t), t),
\end{align}
where $x \in \mathbb{R}^n$ is the state variable, $u \in \mathbb{R}^l$ is the control input vector, and $f \in \mathbb{R}^n$ is a (possibly time-varying) vector field of the state $x(t)$, control input $u(t)$ and the time $t$.

Following the conventional definition of viability theory and viable set \cite{aubin2009viability},  we define \textit{temporal} viability and \textit{time-varying} viable set as follows.

\begin{definition}
(\textbf{Temporal viability and time-varying viable set}) Consider a control system described by a differential equation $\dot x(t) = f(x(t), u(t),t)$  in \eqref{eq:control_system}. A subset $\mathcal{F}(t) \in \mathbb{R}^n$ enjoys the \textit{temporal} viability property for the system $\dot x(t)$ under the time interval $t \in [\tilde t, \bar t]$ if for every initial state $x(\tilde t) \in \mathcal{F}(\tilde t)$ at time $\bar t$, there exists at least one solution to the system
starting at $x(\tilde t)$ which is viable in the time interval $[\tilde t, \bar t]$ in the sense that $$\forall t \in [\tilde t, \bar t], x(t) \in \mathcal{F}(t).$$

The set $\mathcal{F}(t)$ is then termed a \textit{time-varying} viable set for the dynamical control system \eqref{eq:control_system}.
\end{definition}

In the following, without loss of generality we will assume the initial time $\tilde t = 0$. The time $\bar t$ then denotes the maximum existence time that extends the solution of the dynamical system  \eqref{eq:control_system}.  If the solution of the dynamical system \eqref{eq:control_system} can be extended to infinity, we may also consider all the positive time $\bar t \rightarrow \infty$. When a differential equation that models a dynamical control system involves discontinuous right-hand side (e.g., switching controls), we understand its solutions in the sense of Filippov \cite{cortes2008discontinuous}.

We define a distance function for a point $y$ to a (possibly time-varying) set $\mathcal{F}(t)$ as $d^t(y, \mathcal{F}(t))=:  \inf\limits_{z(t) \in \mathcal{F}(t)} \|y-z(t)\|$ at time $t$.  Following \cite{aubin2009set} and \cite{hauswirth2018time}, one can define the temporal tangent cone as follows.

\begin{definition}
(\textbf{Temporal contingent cone}) Let $\mathcal{F}(t)$ be a nonempty subset of $\mathcal{X} \in \mathbb{R}^n$ and $x(t)$ belongs to $\mathcal{F}(t)$ at time $t = \hat t$. The temporal
contingent cone \footnote{We remark that, as has been shown in \cite{hauswirth2018time}, the temporal contingent cone is not necessarily a cone. Similarly to \cite{hauswirth2018time},
we follow the convention and term it `temporal contingent cone' as it reduces to the standard contingent cone in Definition \ref{def:contingent_cone} when the set $\mathcal{F}$ is time independent.} to $\mathcal{F}(t)$ at $x(t)$ and time $\hat t$ is the set
\begin{align}
T_\mathcal{F}^{\hat t}(x) = \left\{v \in \mathcal{X} |\,\,\,\,  \liminf\limits_{h \rightarrow 0^+} \frac{d^{\hat t}\left(x +hv, \mathcal{F}(\hat t +h)\right)}{h} =0  \right\}. %_{h \right 0^+}  % \frac{d_K(x +hv)}{h} =0
\end{align}
\end{definition}

% \{Note I: it has been shown in \cite{hauswirth2018time} that the temporal contingent cone is not necessarily a cone---probably we may change it to a different name...\}

% \{Note II: In the case that the set $\mathcal{F}$ is  time-invariant,  the contingent cone is irrelevant to the norm used in the definition of a point-set distance function. I think this also holds true for the case of temporal contingent cone---but we may need a proof to show this. Otherwise we can simply use the standard Euclidean 2-norm.\}

Now consider a (possibly time-varying) set $\mathcal{F}$ parameterized by an inequality constraint of a time-varying real vector function $g(x(t), t): \mathbb{R}^n \times \mathbb{R} \rightarrow \mathbb{R}^m$:
\begin{align} \label{eq:parameterize_set_F}
    \mathcal{F}(t) = \{(x(t),t) | g(x(t), t) \leq 0\}.
\end{align}

We impose the following assumption on the vector function $g$ for deriving a well-defined temporal contingent cone with favourable properties.
\begin{assumption} \label{assum:function_g}
The function   $g: \mathbb{R}^n \times \mathbb{R} \rightarrow \mathbb{R}^m$ is a $C^1$ (i.e., continuously differential) function of the state $x(t)$, and is Lipschitz continuous with respect to  the time $t$.
\end{assumption}

The following result characterizes an explicit formula of temporal contingent cone $T_\mathcal{F}^{t}(x)$ when the set $\mathcal{F}(t)$ is parameterized by   time-varying functions $g(x(t), t)$ as in \eqref{eq:parameterize_set_F} under Assumption \ref{assum:function_g}.
\begin{lemma} \label{lemma_temporal_cone}
Consider the time-varying set $\mathcal{F}(t)$ parameterized by  time-varying functions $g(x(t), t)$ in \eqref{eq:parameterize_set_F}. Assume that the gradient $\nabla_x g(x(t), t)$ is of full rank for the points $x(t)$ with $g(x(t), t) = 0$. Then the temporal contingent cone is described by
\begin{align}
    T_\mathcal{F}^{t}(x) = \mathbb{R}^n, \,\,\,\,\forall \{(x(t), t)|  g(x(t), t) <0\},
\end{align}
% and
% \begin{align} \label{eq:active_set_cone}
%     T_\mathcal{F}^{t}(x) = \left\{v \middle| \begin{bmatrix}
%      \nabla_x g(x(t), t) \\
%       \nabla_t g(x(t), t)
%     \end{bmatrix}^T
%     \begin{bmatrix}
%      v \\
%       1
%     \end{bmatrix} \leq 0 \right\} \nonumber \\
%     \,\,\,\,\forall \{(x(t), t)|  g(x(t), t) =0\}
% \end{align}

% \begin{align}
%     T_\mathcal{F}^{t}(x) = \mathbb{R}^n, \,\,\,\,\forall \{(x(t), t)|  g(x(t), t) <0\}
% \end{align}
and
\begin{align} \label{eq:active_set_cone}
    T_\mathcal{F}^{t}(x) = \left\{v \middle| \begin{bmatrix}
     (\nabla_x g(x(t), t))^T,
      \nabla_t g(x(t), t)
    \end{bmatrix}
    \begin{bmatrix}
     v \\
      1
    \end{bmatrix} \leq 0 \right\}, \nonumber \\
    \,\,\,\,\forall \{(x(t), t)|  g(x(t), t) =0\}.
\end{align}

\end{lemma}
A detailed calculation of the contingent cone and the proof is omitted here. The case with active constraint (the set under \eqref{eq:active_set_cone}) has also been discussed in \cite{hauswirth2018time}. If a subset of the inequality constraint of the vector function $g$ becomes active, then the formula \eqref{eq:active_set_cone} in Lemma \ref{lemma_temporal_cone} applies to only the subset of active inequality constraints.

\begin{remark}
  A non-empty temporal contingent cone $T_\mathcal{F}^{t}(x)$   for all time $t$ is a necessary condition to ensure the existence of the control input $u(t)$ associated with the time-varying vector field $f(x(t), u(t), t)$. A sufficient condition to guarantee non-empty $T_\mathcal{F}^{t}(x)$ along the solution $x(t)$ and time $t$ is the \textit{forward Lipschitz continuity} of the set $\mathcal{F}(t)$ with respect to time $t$ (see \cite[Theorem 1]{hauswirth2018time}). According to \cite[Proposition 4]{hauswirth2018time}, a sufficient condition to ensure the forward Lipschitz continuity of the set $\mathcal{F}(t)$ is (i) the gradient vector $\nabla_x g(x(t), t)$ has full rank, and (ii) the time-varying function $g(x(t, t))$ is Lipschitz continuous in $t$. In this paper we may further suppose $g(x(t), t)$ is a $C^1$ function of both the state $x(t)$ and time $t$, \footnote{Extensions to piece-wise $C^1$ functions are also possible, which we retain for future research. } which automatically guarantees the second condition. By imposing Assumption \ref{assum:function_g}, in the following we always ensure  that the set $\mathcal{F}(x(t),t)$ parameterized by a set of  inequality/equality constraints of the time-varying function $g(x(t), t)$ is forward Lipschitz continuous.
\end{remark}

Now we present the first main result of this paper.

\begin{theorem} \label{theorem:invariant_temporal_set}
(\textbf{Controlled temporal invariant set})  Consider a forward Lipschitz continuous time-varying set $\mathcal{F}(t)$ parameterized by an inequality constraint of   time-varying  functions $g(x(t), t)$: $\mathcal{F}(t) = \{(x, t)| g(x(t), t) \leq 0\}$. Then the set $\mathcal{F}(t)$ is positively temporal invariant under the dynamical control system $\dot x(t)$, $t \in [0, \bar t]$, of \eqref{eq:control_system} if $x(0) \in \mathcal{F}(0)$, and $\dot x(t) =  f(x(t), u(t), t) \in T_\mathcal{F}^{t}(x)$ with the temporal contingent cone $T_\mathcal{F}^{t}(x)$ derived in Lemma \ref{lemma_temporal_cone}. Equivalently, to guarantee the controlled temporal invariance of the set $\mathcal{F}(t)$, the (possibly time-varying) vector function $f$ should satisfy
\begin{align}
f(x(t), u(t), t) \in \mathbb{R}^n, \,\,\, \forall x(t):  g(x(t), t) < 0, \forall t \in [0, \bar t],
\end{align}
or
\begin{align} \label{eq:control_input_general}
\nabla_x g(x(t), t) ^T f(x(t), u(t), t) + \nabla_t g(x(t), t) \leq 0, \nonumber \\ \,\,\, \forall x(t):  g(x(t), t) = 0, \forall t \in [0, \bar t].
\end{align}
\end{theorem}
\textbf{Proof:}
%\begin{proof}
The proof is based on the explicit formulas in Lemma \ref{lemma_temporal_cone} that characterizes the set of temporal contingent cone along time. Forward Lipschitz continuity of the set $\mathcal{F}(t)$, which is guaranteed by Assumption \ref{assum:function_g} on the constraint function  $g(x(t), t)$ and full rank of $\nabla_x g(x(t), t)$, implies a non-empty set of temporal contingent cone:
\begin{align}
    T_\mathcal{F}^{t}(x) \neq \emptyset, \forall t \in [0, \bar t].
\end{align}
The necessary and sufficient condition to ensure that the time-varying set $\mathcal{F}(t)$ is viable under the time-varying control system $\dot x(t) =  f(x(t), u(t), t)$ is
\begin{align} \label{eq:necessary_suffi_condition}
    f(x(t), u(t), t) \cap T_\mathcal{F}^{t}(x)  \neq \emptyset, \forall t \in [0, \bar t].
\end{align}
When the time-varying constraint functions are inactive in the sense that $g(x(t), t) < 0$, the temporal contingent cone is the whole space  $T_\mathcal{F}^{t}(x) = \mathbb{R}^n$, which implies that the time-varying vector function can be any vector $f(x(t), u(t), t) \in \mathbb{R}^n$.
When $g(x(t), t) = 0$  at which constraint functions become active, the temporal contingent cone formula in \eqref{eq:active_set_cone} renders an equivalent formulation as in  \eqref{eq:control_input_general} to ensure that the viability condition of  \eqref{eq:necessary_suffi_condition} is always satisfied.  The control input \eqref{eq:control_input_general} serves  corrective actions that regulate the states of the dynamical system $\dot x$ to be controlled temporal  invariant in the set $\mathcal{F}(t)$.   %\qed

%\end{proof}
The above theorem extends the classical Nagumo theorem and  standard results in controlled invariance theory (see e.g., \cite{blanchini1999set}).

\section{Temporal viability regulation for control affine systems}  \label{sec:control_affine_system}
In this section we will focus on the control affine system described by the following general form
\begin{align} \label{eq:control_affine_system}
    \dot p(t) = f_0(p(t)) + \sum_{j = 1}^l f_j(p(t)) u_j(t),
\end{align}
where $p \in \mathbb{R}^{n}$ is the system state, $f_{0}$ is a smooth drift function term,   $u_{j}$ is the \emph{scalar} control input (possibly time-varying) associated with the smooth vector field $f_{j}$, and $l$ is the number of vector field functions.
%Note that the set of vector fields $f_{i,j}$, $j = 0,1, \cdots, l$ can be thought of defining a subspace of the tangent plan at each point.
Such a nonlinear control affine system~\eqref{eq:control_affine_system} with a drift term is very general in that it describes many different types of real-life vehicle dynamics and control systems, including control systems subject to under-actuation or nonholonomic motion constraints, as we will show in the next section. We remark that a  nonlinear control affine system \eqref{eq:control_affine_system} with drifts can be equivalently described by the following \textit{affine} distribution (see e.g., \cite{isidori1995nonlinear} and \cite{nijmeijer1990nonlinear})
\begin{align}
\Delta  = f_{0} + \text{span}\{f_{1}, f_{2}, \cdots, f_{l}\}.
\end{align}

%\subsection{Temporal viability regulation}
When specializing the temporal viability theory to  control affine systems \eqref{eq:control_affine_system}, one obtains the following theorem on temporal viability regulation.
\begin{theorem} \label{theorem:invariant_temporal_set_affine}
(\textbf{Temporal viability regulation})  Consider a forward Lipschitz continuous time-varying set $\mathcal{F}(t)$ parameterized by an inequality constraint of   time-varying  functions $g(p(t), t)$: $\mathcal{F}(t) = \{(p, t)| g(p, t) \leq 0\}$. The set $\mathcal{F}(t)$ is controlled temporal viable under the control affine system $\dot p(t)$, $t \in [0, \bar t]$, of \eqref{eq:control_affine_system} if $p(0) \in \mathcal{F}(0)$ and the control input $u_j$ satisfies  (whenever the inequality $g(p(t), t) = 0$ is active):
\begin{align} \label{eq:viable_condition_affine}
 \sum_{j = 1}^l u_j &  \nabla_p g(p(t), t) ^T  f_j(p(t)) \nonumber \\ & \leq  -\nabla_p g(p(t), t) ^T f_0(p(t))    -\nabla_t g(p(t), t) , \nonumber \\ \,\,\, &\forall p(t):  g(p(t), t) = 0, \forall t \in [0, \bar t];
\end{align}
and when the inequality constraint is inactive the control input takes arbitrary value in the sense that $u_j \in \mathbb{R}, j = 1, \cdots, l$.
\end{theorem}

The above theorem can be obtained as a consequence of Theorem \ref{theorem:invariant_temporal_set}, and a proof is omitted for space consideration.

\subsection{Special Case: Viability Regulation under Time-varying Equality Constraints}
Now we consider a special case that the  $\mathcal{F}(t)$ is parameterized by time-varying \textit{equality} constraints:
\begin{align} \label{eq:parameterize_set_F_equality}
    \mathcal{F}(t) = \{(p(t), t) | g(p(t), t) = 0\},
\end{align}
where $g: \mathbb{R}^n \times \mathbb{R} \rightarrow \mathbb{R}^m$ is a $C^1$ vector function of the state $p(t)$ and is Lipschitz continuous with respect to the time $t$.
Following Theorem \ref{theorem:invariant_temporal_set_affine} we have the following result.

\begin{lemma}
Consider a forward Lipschitz continuous time-varying set $\mathcal{F}(t)$ parameterized by time-varying equality constraint in \eqref{eq:parameterize_set_F_equality}. The set $\mathcal{F}(t)$ is controlled temporal viable under the control affine system $\dot p(t)$ of \eqref{eq:control_affine_system} if $p(0) \in \mathcal{F}(0)$ and the control input $u_j$ satisfies
\begin{align} \label{eq:viable_condition_affine_equality}
&\sum_{j = 1}^l u_j \nabla_p g(p(t), t) ^T f_j(p(t)) \nonumber \\ &=  -\nabla_p g(p(t), t) ^T f_0(p(t)) -\nabla_t g(p(t), t).
\end{align}
\end{lemma}
The above lemma recovers the main result in a recent paper \cite{Invariance2018MTNS}.
%\subsection{Controlled invariant set parameterized by multiple constraint functions}
% We impose the following assumption on the vector field functions.
% \begin{assumption} \label{assum_full_rank_distribution}
% The distribution $\text{span}\{f_{1}, f_{2}, \cdots, f_{l}\}$
% %$\text{span}(p(t))\{f_{1}(p(t), f_{2}(p(t)), \cdots, f_{l}(p(t))\}$
% obtained from the $l$ vector field functions $f_1(p(t))$,  $f_2(p(t))$,  $\cdots, f_l(p(t))$ is of full row rank at the point $p = p(t)$.
% \end{assumption}
%\subsection{Existence of control input}
In the case of time-varying equality constraint, the control input condition in  \eqref{eq:viable_condition_affine_equality} should satisfy a much more restrictive condition (in terms of equality) than that in the case of time-varying set parameterized by inequality constraint.
The following lemma presents a sufficient condition to ensure the existence of the control input in  \eqref{eq:viable_condition_affine_equality}.
\begin{lemma}
Suppose that the set $\mathcal{F}(t)$ is forward Lipschitz continuous, and that the distribution $\text{span}\{f_{1}, f_{2}, \cdots, f_{l}\}$
%$\text{span}(p(t))\{f_{1}(p(t), f_{2}(p(t)), \cdots, f_{l}(p(t))\}$
obtained from the $l$ vector field functions $f_1(p(t))$,  $f_2(p(t))$,  $\cdots, f_l(p(t))$ is of full row rank at the point $p = p(t)$.  Then the control inputs $u_j$ always exist for \eqref{eq:viable_condition_affine} that guarantees temporal viability regulation.
\end{lemma}

\textbf{Proof:}
%\begin{proof}
Forward Lipschitz continuity of the set $\mathcal{F}(t)$ implies the full rank of the gradient vector $\nabla_p g(p(t), t)$, which further implies that the map $\nabla_p g(p(t), t)^T$ is surjective. Denote $\mathcal{M}: = \nabla_p g(p(t), t)^T$, and rewrite the equality condition from \eqref{eq:viable_condition_affine_equality} as follows
\begin{align} \label{eq:viable_condition_affine_equality_rewritten}
\mathcal{M} \sum_{j = 1}^l   f_j(p(t)) u_j =  -\mathcal{M} f_0(p(t)) -\nabla_t g(p(t), t).
\end{align}
Sujectivity of $\mathcal{M}$ implies the existence of solutions to the above equality condition (or equivalently, the equation in \eqref{eq:viable_condition_affine_equality}), which indicates that the set parameterized by $\sum_{j = 1}^l    f_j(p(t)) u_j$ is non-empty. The second condition indicates that the distribution map $\text{span}\{f_{1}, f_{2}, \cdots, f_{l}\}$ is surjective, which then guarantees the existence of the control input $u_j$. %\qed
%\end{proof}

The above lemma only shows a sufficient yet strong condition for guaranteeing existence of viable controls. Other weaker sufficient conditions also exist, and a detailed characterization of viable control input will be in the future research. In practice, analytical solutions may be hard to obtain since the solving of \eqref{eq:viable_condition_affine_equality} involves algebraic equations of both system states and the time. In this way, available symbolic toolboxes and computational approaches (see e.g., \cite{kwatny2000nonlinear}) can be used to find admissible and viable control input for a specified dynamical system. The following example will be treated using symbolical tools, but the reader is referred to~\cite{sun2018feasibility} for a heuristic solving algorithm and a set of illustrative and analytically computed solutions for the time-invariant case.

\section{Application Study: Vehicle Coordination with time-varying motion constraints}
\label{sec:applications}
In this section, we apply the developed theory on temporal viability regulation to coordinate two unicycle vehicles, subject to time-varying equality and inequality motion constraints on their joint state. First, we give the relevant vehicle models, whose kinematics can be equivalently  formulated as (A) time-invariant kinematic equality constraints. We then proceed to detail (B) time-varying velocity equality constraints, (C) time-varying inequality constraints on the inter-vehicle distance and (D) a time-varying visibility inequality constraint. Finally, we present a closed loop simulation where the system controls are regulated in accordance with Theorem~\ref{theorem:invariant_temporal_set_affine} subject to (A)-(D), thereby yielding a state trajectory which satisfies all constraints at all times $t\in[0,\bar{t}]$.

\subsection{Vehicle Models}
Consider a unicycle model, described by the equations
\begin{align} \label{eq:example_unicycle}
\dot x_i &= v_i \,\, \text{cos}(\theta_i), \nonumber \\
\dot y_i &= v_i \,\, \text{sin}(\theta_i),   \\
\dot \theta_i &=   u_i,  \nonumber
\end{align}
where the state of system $i$ is configured on $[x_i, y_i, \theta_i]^{\top} \in \mathbb{R}^2 \times \mathbb{S}^1 \in \mathbb{R}^3$.
This kinematic equation can be equivalently stated by the annihilating codistribution
\begin{align}\label{eq:annihilatingcodistributionKin}
\Omega_{K,i} := \text{sin}(\theta_i) \text{d}x_i-\text{cos}(\theta_i) \text{d}y_i.
\end{align}
In this example, we consider coordination of $n$ such vehicles, whose joint states are denoted $p(t)\in \mathbb{R}^{2n} \times \mathbb{S}^n$, where then
\begin{align}\label{eq:kinematics}
\Omega_{K,i} \dot{p} = 0,\qquad \forall i\in 0,\cdots, n.
\end{align}

\subsection{Time-varying Velocity Equality Constraints}
In the context of temporal viability regulation, it may be useful to restrict certain parts of the joint state to desired state trajectories. To see how, let $v_{i,r}(t), u_{i,r}(t)\in C^0$ define a desired reference trajectory of the $i^{th}$ vehicle, which, similar to~\eqref{eq:annihilatingcodistributionKin}, may be expressed by the annihilating codistribution
$$
\cos(\theta_i)\text{d}x_i + \sin(\theta_i)\text{d}y_i = v_{i,r}(t), \quad \text{d}\theta_i = u_{1,r}(t),
$$
here written compactly with $\Omega_{E,i}\dot{p} = T_{E,i}$, where then
\begin{equation}\label{eq:timevaryingVelocity}
\Omega_{E,i}\hspace{-1pt}=\hspace{-1pt}
\begin{bmatrix}
\cos(\theta_i)\text{d}x_i \hspace{-1pt}+\hspace{-1pt} \sin(\theta_i)\text{d}y_i \\
\text{d}\theta_i
\end{bmatrix}
,\;
T_{E,i}\hspace{-1pt}=\hspace{-1pt}
\begin{bmatrix}
v_{i,r}(t)\\
u_{i,r}(t)
\end{bmatrix}.
\end{equation}

\subsection{Time-Varying Distance Inequality Constraints}
In mobile vehicle coordination tasks, it is often useful to pose inequality constraints on inter-vehicular distance. Let
$$
\frac{1}{2}d_{ij}^{-}(t)^2\leq \frac{1}{2}(x_i - x_j)^2 +  \frac{1}{2}(y_i - y_j)^2  \leq \frac{1}{2}d_{ij}^{+}(t)^2
$$
for some time-varying lower and upper bounds $0\leq d_{ij}^{-}(t)<d_{ij}^{+}(t)$, which may be expressed as a vector valued inequality
$$
g_{ij}^d(p, t) := \frac{1}{2}
\begin{bmatrix}
+(x_i - x_j)^2 +  (y_i - y_j)^2 - d_{ij}^{+}(t)^2\\
-(x_j - x_i)^2 -  (y_i - y_j)^2 + d_{ij}^{-}(t)^2\\
\end{bmatrix} \leq 0.
$$
The utility of such a constraint cannot be under-stated, as the time-varying constraints may be dynamically updated when interacting with the environment with $d^-_{ij}(t)$ and $d^{+}_{ij}$ being  $C^1$ functions of time.
The constraint can be written on the form of Theorem 3, as $\Omega_I^d \dot{p} \leq T_I^d$ where
\begin{align}
\Omega_{I,ij}^d &:= \nabla_p g_{ij}^d(p, t) \nonumber \\
&=
\begin{bmatrix}
(x_i-x_j)(\text{d}x_i-\text{d}x_j) + (y_i-y_j)(\text{d}y_i-\text{d}y_j)
\\
(x_j-x_i)(\text{d}x_i-\text{d}x_j)  + (y_j-y_i)(\text{d}y_i-\text{d}y_j)
\end{bmatrix},  \nonumber \\ \label{eq:timevaryingDistance}
T_{I,ij}^d &:= -\nabla_t g_{ij}^d(p, t)=
\begin{bmatrix}
+d_{ij}^{+}(t)\dot{d}_{ij}^{+}(t)
\\
-d_{ij}^{-}(t)\dot{d}_{ij}^{-}(t)
\end{bmatrix}.
\end{align}
where we have used the standard dual bases $\{\text{d}x_i, \text{d}y_i, \text{d}\theta_i, \text{d}x_j, \text{d}y_j, \text{d}\theta_j\}$ in representing the annihilating codistribution associated with the gradient vector.
\subsection{Time-Varying Visibility Inequality Constraints}
\begin{figure}[t]
\begin{center}
\vspace{-10pt}
\includegraphics[width=0.5\textwidth]{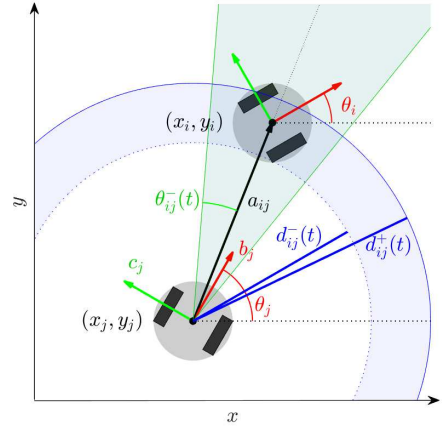}
\vspace{-10pt}
\caption{Illustration of a visibility inequality constraint, ${g}^{v}_{ij}(p,t)$, bounding the direction $b_j$ to the green cone defined by $a_{ij}$ and the half apex angle $\theta_{ij}^-$, and the two-sided distance inequality constraint (blue).}
\label{fig:geometry}
\end{center}
\end{figure}

The second class of inequality constraint under consideration is a visibility constraint posed on the relative rotation of vehicles, presented in a time-invariant form in~\cite{sun2018feasibility}. Here, the cosine angle of the body direction of system $j$, $b_{j} := [\cos(\theta_j), \sin(\theta_j)]$, and the the direction of system $i$ relative to $j$, $a_{ij} := [x_i - x_j, y_i - y_j]$ is bounded (see Figure~\ref{fig:geometry}). In other words, we enforce a constraint on the time-varying apex angle, $2\theta_{ij}^-(t)$, defining a cone of visibility of system $j$, such that
\begin{align*}
g_{ij}^v(p, t) := \cos(\theta_{ij}^-(t)) - \frac{\langle a_{ij},b_{j} \rangle}{\langle a_{ij}, a_{ij} \rangle^{1/2}}\leq  0.
\end{align*}
By letting $c_{j} := [-\sin(\theta_j), \cos(\theta_j)]$, the associated annihilating codistribution of this inequality constraint becomes
\begin{align}
\Omega_{I,ij}^v\hspace{-1pt}:=&\nabla_p g_{ij}^v(p, t) \nonumber \\
=&\hspace{-1pt}\dfrac{\langle a_{ij}, c_j\rangle}{\sqrt{\langle a_{ij}, a_{ij} \rangle}} \Bigg(
\dfrac{1}{\langle a_{ij}, a_{ij} \rangle} \Bigg\langle a_{ij},
\hspace{-3pt}
\Bigg[
\begin{array}{c}
\hspace{-5pt}\text{d}x_i-\text{d}x_j\hspace{-5pt}\\
\hspace{-5pt}\text{d}y_j-\text{d}y_i\hspace{-5pt}
\end{array}
\Bigg]
\hspace{-1pt}
\Bigg\rangle
\hspace{-2pt}+
\hspace{-2pt}\text{d}\theta_j\Bigg).\nonumber
\\\label{eq:timevaryingAngle}
T_{I,ij}^v :=& -\nabla_t g_{ij}^v(p, t) = \sin(\theta^-_{ij}(t))\dot{\theta}^-_{ij}(t).
\end{align}
With these general descriptions of the annihilating codistributions associated with the time-varying constraints, we proceed to show how they may be implemented in practice.

\subsection{Coordination with Time-varying Constraints}
We now tackle the challenging task of controlling a system consisting of $n=2$ unicycle vehicles subject to constraints on the form (A)-(D) posed on the joint state of the vehicles, $p(t)\in \mathbb{R}^4 \times \mathbb{S}^2$. We use the following seven constraints
\begin{align*}
\Omega_{K,i}\dot{p}&=0,\;\; i=1,2,
\hspace{19.5pt}(\text{see equation}~\eqref{eq:kinematics})\\
\Omega_{E,1}\dot{p}&=T_{E,1},
\hspace{47pt}(\text{see equation}~\eqref{eq:timevaryingVelocity})\\
\Omega_{I,12}^d\dot{p}&\leq T_{I,12}^d,
\hspace{45pt}(\text{see equation}~\eqref{eq:timevaryingDistance})\\
\Omega_{I,12}^v\dot{p}&\leq T_{I,12}^v, \hspace{45pt}(\text{see equation}~\eqref{eq:timevaryingAngle})
\end{align*}
with the right hand side defined by arbitrary five time-varying functions, here for illustrative purpose chosen as
\begin{subequations}
\begin{align}
v_{1r}(t)&=|2\sin(2t)|,\\ u_{1r}(t)&={0.7}^{-1}v_{1r}(t)\tan(2\sin(t)\cos(t)),\\
d^+_{12}(t)&= 1.2+0.1\sin(5t),\\
d^-_{12}(t)&= 0.9-0.1\sin(t),\\
\theta^-_{12}(t) &= 0.1+0.05\sin(2t).
\end{align}
\end{subequations}
In effect, this set of constraints result in both vehicles satisfying the kinematic equations of a unicycle in equation~\eqref{eq:example_unicycle}, while  vehicle $i=1$ follows a path given by $v_{1r}(t), u_{1r}(t)$, and vehicle $i=2$ maintains a set of time-varying inequality constraints involving distance and visibility with respect to the vehicle $i=1$.

For every possible combination of active inequality constraints, we can solve the resulting set of algebraic equations in Theorem~\ref{theorem:invariant_temporal_set_affine} offline in a symbolical sense. In our case, there is a total of three inequality constraints, generating eight combinations of possible active constraints. Each of these solutions will be associated with varying degrees of freedom, $l=0,1,2$, in which we may choose any $U(t)\in \mathbb{R}^l$ with elements $\mu_j(t)$ (solutions to the derived algebraic inequality) such that Theorem~\ref{theorem:invariant_temporal_set_affine} is satisfied at a time $t$. In the simulation, we restrict $U(t)\in [-3,3]^l$ and simply pick the controls which minimizes $\|U(t)\|_{\infty}$ subject to the active algebraic constraints. These virtual control inputs may instead be selected to minimize a cost on the state or achieve some sense of robustness, but this is left for future research.

\begin{figure}[t]
\begin{center}
\vspace{-10pt}
\includegraphics[width=0.5\textwidth]{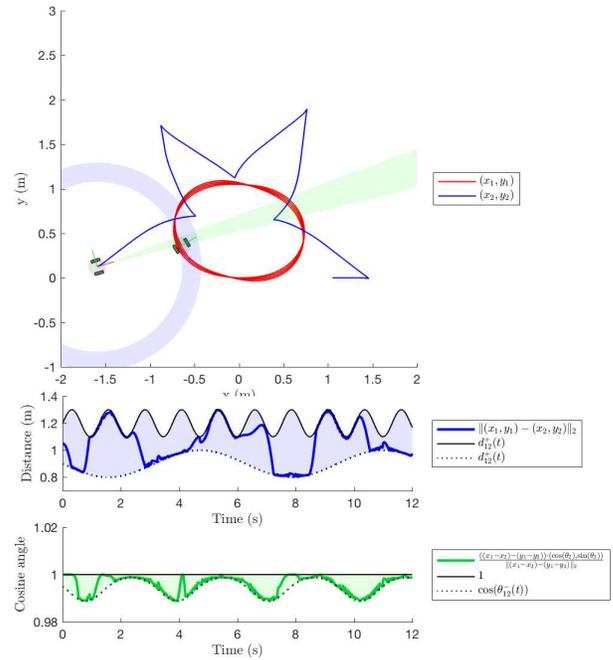}
\vspace{-30pt}
\caption{Simulation with two vehicles, controlled by temporal viability regulation. \textit{Top:} Positional state trajectories of the vehicles, $i=1$ (red) and $i=2$ (blue), with the inequality constraints in distance (blue) and visibility (green) at the time $t=12$. \textit{Center:} The time-varying distance inequality constraint (center). \textit{Bottom:} The narrow time-varying visibility constraint. Simulation video is available upon request. }
\label{fig:simulationresult}
\end{center}
\end{figure}

\begin{figure}[t]
\begin{center}
\includegraphics[width=0.5\textwidth]{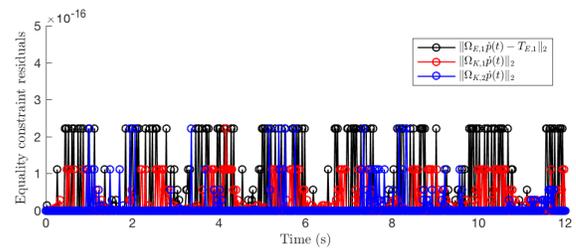}
\caption{Residuals of the equality constraints  in the two-norm in time, with the kinematic constraint residuals (blue,red) and the time-varying velocity constraint residuals (black).}
\label{fig:simulationresultB}
\end{center}
\end{figure}

When studying the simulation result in Figure~\ref{fig:simulationresult}, it is clear that the time-varying inequality constraints on the inter-vehicular distance and the visibility are respected at all times. Note that the choice of $\theta_{12}^-(t)$ generates a very narrow  cone of visibility at certain points in time, and that one or more of the inequality constraints are saturated at almost all times. To verify that the kinematic equality constraints and time-varying velocity inequality constraints are met at all times, the residuals of these constraints are shown in the 2-norm in time (see Figure~\ref{fig:simulationresultB}). Clearly, the computed state trajectory of the joint state satisfy all posed equality constraints to the numerical precision of Matlab, at $2.22\cdot 10^{-16}$.

\section{Conclusions}\label{sec:conclusions}
In this paper, we discuss temporal viability regulation for general dynamical control systems with a particular focus on control affine systems. The aim is to address typical control scenarios and provide a constructive approach to enforcing a time-varying set of equalities and/or inequalities on the state. We present  control laws for ensuring temporal viability and controlled invariance for arbitrary control affine systems with time-varying viable set constraints, and illustrate the proposed theory by an example with homogeneous mobile vehicle coordination under time-varying motion constraints.

\section*{Acknowledgement}
The authors would like to thank Prof. Daniel Zelazo for suggesting the paper \cite{hauswirth2018time}, and for his insightful discussions on this topic.

\bibliography{main}
\bibliographystyle{ieeetr}

 \iffalse
\begin{tikzpicture}[scale = 0.5]
\pgfmathsetmacro{\dW}{1.0}
\pgfmathsetmacro{\cW}{0.6}
\pgfmathsetmacro{\hW}{1.2}
\begin{scope}[shift={(3,5)},rotate=0]
\draw[fill=black!20!white] (0,0) circle (2);
\draw[thick,fill=black!80!white] (\cW+\dW,-\hW) -- (\cW+\dW,\hW) -- (\dW,\hW) -- (\dW,-\hW) -- (\cW+\dW,-\hW);
\draw[thick,fill=black!80!white] (-\cW-\dW,-\hW) -- (-\cW-\dW,\hW) -- (-\dW,\hW) -- (-\dW,-\hW) -- (-\cW-\dW,-\hW);
\draw[->,>=stealth,ultra thick, red] (0,0) -- (0,3);
\draw[->,>=stealth,ultra thick, green!80!black] (0,0) -- (-3,0);
\end{scope}
\end{tikzpicture}
\fi

\end{document}